\documentclass[twocolumn,showpacs,aps,prl,amsmath,amssymb,citeautoscript,longbibliography,floatfix,superscriptaddress,reprint]{revtex4-1}

\pdfoutput=1
\usepackage{graphicx}
\usepackage[utf8]{inputenc}
\usepackage[T1]{fontenc}
\usepackage{placeins}
\usepackage{subfiles}
\usepackage{amsmath,amsthm,amsbsy,amssymb,braket,mathtools,bbm}
\usepackage[english]{babel}
\usepackage[caption=false]{subfig}
\usepackage{blindtext}
\usepackage[colorlinks, citecolor=blue, linkcolor=blue, urlcolor=blue]{hyperref}
\usepackage{etoolbox}
\usepackage{paralist}
\newcommand{\nn}				{\nonumber}

\newcommand{\E}	[1]		        {\mathrm{e}^{#1}}

\newcommand{\D}               {\mathrm{d}}
\newcommand{\Hamil}               {\hat{\mathcal H}}

\newcommand{\bsy} [1]              {\boldsymbol{#1}}
\newcommand{\abs} [1]              {\vert #1 \vert}
\newcommand{\p} [1]              {\bsy{\sigma}_{#1}}

\DeclareMathOperator\sign{sgn}

\begin{document}
\title{Creation of spin-triplet Cooper pairs in the absence of magnetic ordering}

\newcommand{\wuerzburg}{Institute for Theoretical Physics and Astrophysics,
	University of W\"{u}rzburg, D-97074 W\"{u}rzburg, Germany}
\newcommand{\aalto}{Department of Applied Physics,
	Aalto University, FIN-00076 Aalto, Finland}

\author{Daniel Breunig}
\affiliation{\wuerzburg}

\author{Pablo Burset}
\affiliation{\aalto}

\author{Bj\"orn Trauzettel}
\affiliation{\wuerzburg}

\date{\today}

\begin{abstract}
In superconducting spintronics, it is essential to generate spin-triplet Cooper pairs on demand. Up to now, proposals to do so concentrate on hybrid structures in which a superconductor (SC) is combined with a magnetically ordered material (or an external magnetic field). 
We, instead, identify a novel way to create and isolate spin-triplet Cooper pairs in the absence of any magnetic ordering. 
This achievement is only possible because we drive a system with strong spin-orbit interaction--the Dirac surface states of a strong topological insulator (TI)--out of equilibrium. In particular, we consider a bipolar TI-SC-TI junction, where the electrochemical potentials in the outer leads differ in their overall sign. As a result, we find that nonlocal singlet pairing across the junction is completely suppressed for any excitation energy. Hence, this junction acts as a perfect spin triplet filter across the SC generating equal-spin Cooper pairs via crossed Andreev reflection. 
\end{abstract}

\maketitle

\begin{figure} 
\renewcommand{\figurename}{Fig.}
\centering
\includegraphics[width=.95\linewidth]{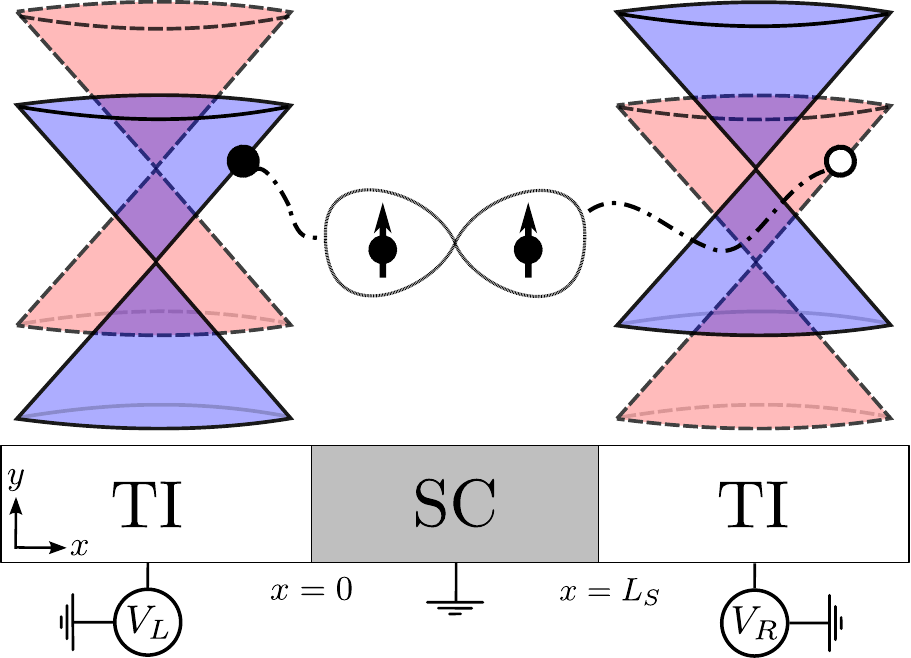}
\caption{Schematic of the TI-SC-TI junction. While the voltages applied to the outer leads should be tunable, the SC is assumed to be grounded. The dispersion relations are depicted for illustration: solid (dashed) lines represent electron (hole) Dirac cones. 
A possible crossed Andreev reflection process--resulting in the injection of equal-spin triplet Cooper pairs in the SC due to spin-momentum locking--is sketched.}
\label{fig:1}
\end{figure}

{\it Introduction.---}In spintronics, it is desirable to achieve spin manipulation in the absence of magnetic fields because electronic switching processes can be done much faster than their magnetic counterparts. The prime example of such a device is the famous Datta-Das transistor based on materials with strong spin-orbit coupling \cite{Datta1990}. 
The situation is similar in superconducting spintronics \cite{Eschrig_2010,SC_Spintronics}. 
However, to the best of our knowledge, no device proposal has been made so far that allows for a performance in the absence of magnetic ordering \cite{[{Proposals for superconductng spin valves require the use of magnetic elements. See, for example, }] [{}]Fominov_2010,*Buechner_2012,*Jiang_2013,*Aarts_2014,*Robinson_2014}. 
Specifically, it would be exciting to create equal-spin Cooper pairs on demand in a device in which spin-orbit coupling is the crucial feature for its application in spintronics. For this vision to come true, the surface states of three-dimensional (3D) topological insulators (TIs) are promising building blocks because they mimic a truly relativistic spin-orbit coupling in a condensed matter setting. We have identified a bipolar device based on two areas of TI surface states connected to each other via a common central superconductor (SC), i.e. a TI-SC-TI junction, that can act as a generator for equal-spin Cooper pairs on demand, {\it cf.} Fig.~\ref{fig:1}. In fact, the working principle of the device is based on crossed Andreev reflection--generated out of equilibrium--which enables a transfer of Cooper pairs (with a finite net spin) from the two TI regions into the SC.

The bipolar TI-SC-TI setup is inspired by a seminal work by Cayssol where a similar junction has been studied in the context of graphene \cite{Cayssol2008}. In that work, the spin degree of freedom played no role because of the weak spin-orbit coupling in graphene. In TIs, instead, strong spin-orbit coupling in combination with superconducting and/or magnetic ordering gives rise to intriguing physics \cite{Linder_2010}, for instance, the emergence of Majorana bound states \cite{Fu_2008,Tanaka_2009} or odd-frequency SC pairing \cite{Yokoyama_2012,Black-Schaffer_2012,Crepin2015,Burset2015}. The underlying reason is that spin-rotational invariance is broken by the spin-orbit coupling and fundamentally different gaps, including a spin triplet state \cite{Tkachov_2013}, can be induced in TI surface states by proximity to an $s$-wave SC and/or a magnetic insulator 
\cite{[{Triplet superconductivity naturally arises in materials or proximitized structures with broken spin-rotation symmetry: }] [{}]Frigeri_2004,*Frigeri_2004b,*Burset_2014,*Maslov_2015,*Bergeret_2015,*Bergeret_2016}. 
From the experimental side, it seems to be more feasible to induce superconducting order into TI surface states than magnetic order, although both tasks have been recently achieved \cite{Wang_2012,Zareapour2012,Oostinga_2012,Maier_2012,Veldhorst_2012,Williams_2012,Mason_2013,Oostinga2013, Sochnikov2015, Wiedenmann2016, Lee2016}. This observation implies that our setup should be directly realizable in the current generation of hybrid structures on TI surfaces.

In this proposal, we assume that an $n$-type region of the Dirac surface states (of a 3D TI) is connected to a $p$-type region via a superconducting electrode, see Fig.~\ref{fig:1} for a schematic. We also want the two TI regions to be biased by separate voltage sources (corresponding to biases $V_L$ and $V_R$ applied to the left ($L$) and the right ($R$) TI, respectively) and the SC to be grounded. 
In such a bipolar TI-SC-TI junction, Andreev reflection in the same lead and electron tunneling between leads can be efficiently turned off, as previously proposed for graphene \cite{Cayssol2008} and ordinary semiconductors \cite{Brinkman_2010}. 
We show below that, for the case of TIs, spin singlet pairing is strongly suppressed, making this device a perfect spin triplet filter where crossed Andreev reflection processes pump equal-spin Cooper pairs into the SC. 

{\it Model.---}To model the TI-SC-TI device, we introduce the basis $[\hat c_\uparrow(\bsy k),\hat c_\downarrow(\bsy k),\hat c^\dagger_\uparrow(-\bsy k),\hat c^\dagger_\downarrow(-\bsy k)]^T$, where $\hat c^\dagger_\sigma(\bsy k)$ is the creation operator of an electron with momentum $\bsy k$ and spin $\sigma\!=\uparrow,\downarrow$. In this basis, we can determine the corresponding Bogoliubov-de Gennes (BdG) Hamiltonian
\begin{align}
\label{eq:1}
\Hamil_{\mathrm{BdG}}=
\begin{pmatrix}
H_0(\bsy k) &i\Delta(x) \p y \\
-i\Delta(x) \p y & -H_0^*(-\bsy k)
\end{pmatrix},
\end{align}
where the electron Bloch Hamiltonian reads as
\begin{align}
\label{eq:2}
H_0(\bsy k)= & v_F\left(\hat k_x\p x+\hat k_y \p y\right)-\mu(x)\p 0  \\ 
\hat = & v_F\left(\hat k_x\p x+k_y \p y\right)-\mu(x)\p 0.
\label{eq:3}
\end{align}
Here, $v_F$ is the Fermi velocity, $\p i$ are the Pauli matrices in spin space, $\hat k_{j}\!\!=\!\!-i\hbar\partial_{j}$ are the momentum operators in position basis, $\Delta(x)$ and $\mu(x)$ define spatially dependent SC pairing and electrochemical potentials, respectively. 
We set $v_F\!=\!\hbar\!=\!1$ in what follows.
In Eq. \eqref{eq:3}, we reduce the Bloch Hamiltonian to a quasi-1D operator, where $k_y$ is now a parameter defining the angle of incidence. This is possible due to the choice of the coordinate system in Fig. \ref{fig:1} and the perfect alignment of the interfaces (between TI and SC) at $x\!=\!0$ and $x\!=\!L_S$ along the $y$-axis.  Experimentally, it is very difficult to resolve the $k_y$--dependence. Hence, we will average subsequent characteristics/observables with respect to $k_y$. 
In our TI-SC-TI-junction, we assume the electrochemical potentials to be constant in each domain. In particular, we define $\mu(x)\!=\!\mu_L \Theta(-x)\!+\!\mu_S\Theta(x)\Theta(L_S\!-\!x)\!+\!\mu_R\Theta(x\!-\!L_S)$, with $\Theta(\cdot)$ the Heaviside function. The superconducting pairing is chosen to be finite only underneath the SC, i.e. $\Delta(x)\!=\!\Delta_0\Theta(x)\Theta(L_S\!-\!x)$. 

Setting up the scattering problem for the TI-SC-TI junction, see the supplemental material (SM \cite{SuppMat}), we obtain the electron and hole wave vectors in the normal leads as $k_{i}^{e/h}\!=\!\zeta_i^{e/h}\abs{\varepsilon\!\pm\!\mu_i}\cos\theta_i^{e/h}$, with $i\!\in\!\{L,R\}$ and $\varepsilon\!>\!0$ the excitation energy.
Here, $\zeta_i^{e/h} \!=\! \sign\left(\varepsilon\!\pm\!\mu_i\!+\!\abs{k_y}\right)$ defines if the particle stems from the valence or the conduction band and $\theta_i^{e/h}\!=\!\arcsin\!\left(k_y/\abs{\varepsilon\!\pm\!\mu_i}\right)$ is the angle of incidence for electrons/holes. 
In such a junction, four transport channels exist: 
\begin{inparaenum}[(i)] 
	\item normal reflection (NR); 
	\item local Andreev reflection (LAR); 
	\item electron cotunneling (CO); and 
	\item crossed Andreev reflection (CAR). 
\end{inparaenum}

\begin{figure} 
\renewcommand{\figurename}{Fig.}
\renewcommand{\thesubfigure}{\roman{subfigure}}
\centering
\includegraphics[width=.95\linewidth]{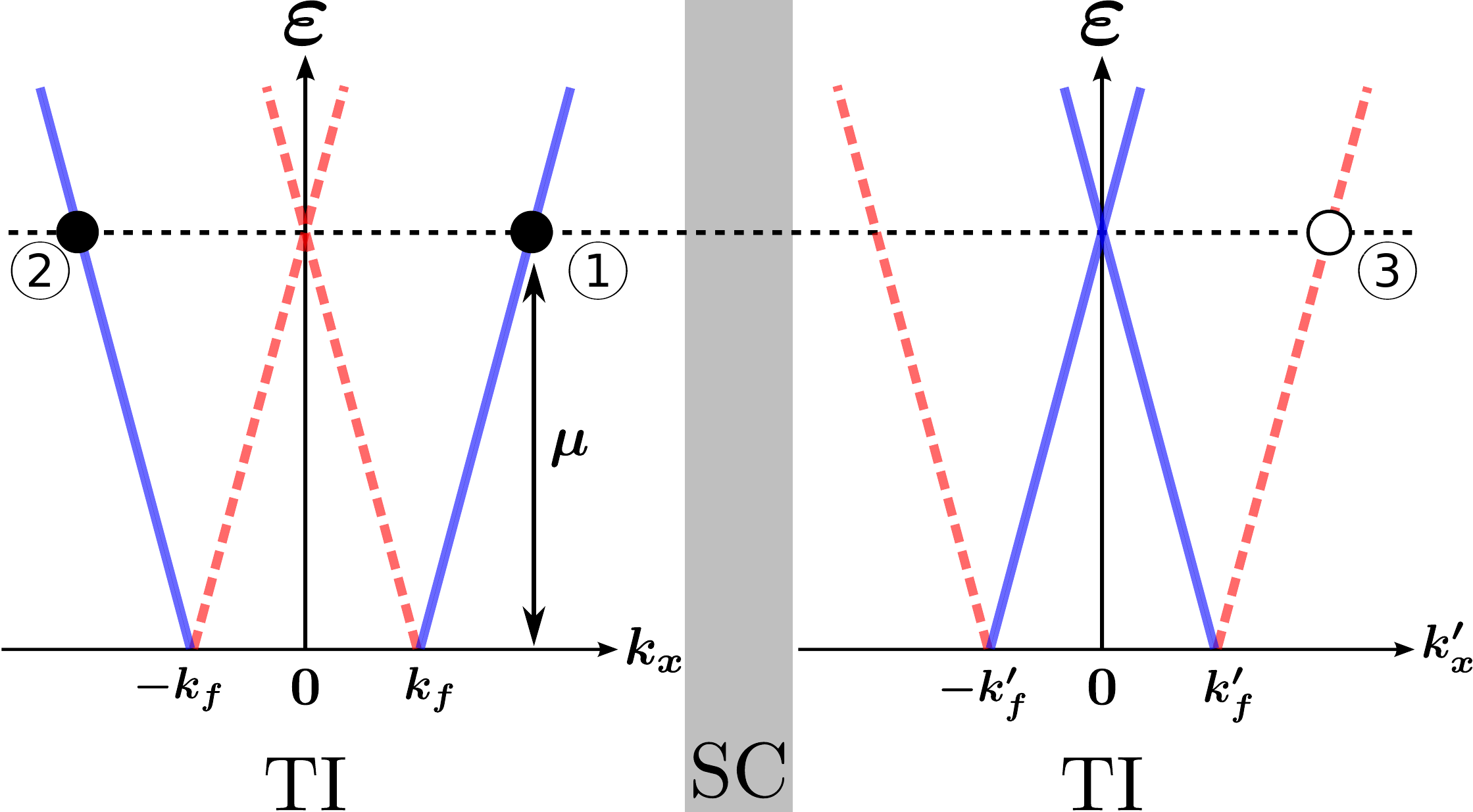}
\caption{Electron (solid lines) and hole (dashed lines) dispersion relations in $L$ and $R$, at $k_y\!=\!0$. The SC is depicted by the gray domain. At $\varepsilon\!=\!\mu$, the incident electron \textbf{1} may be reflected as an electron \textbf{2} at the interface (NR) or transmitted as a hole \textbf{3} through the SC (CAR). LAR and CO are, instead, not permitted since the states at the Dirac points have zero momentum. }
\label{fig:2}
\end{figure}

We obtain a bipolar system by choosing the electrochemical potentials in the TIs to have the same modulus, but different signs, $\mu_L\!=\!-\mu_R\!\equiv\! \mu\!>\!0$. 
Due to the nodal dispersion relation, this choice allows us to completely suppress two out of four transport channels: LAR and CO. 
As illustrated in Fig. \ref{fig:2}, this complete suppression is achieved by applying a voltage $V_L$ to $L$ such that the corresponding excitation energy coincides with the electrochemical potential, i.e. $\mathrm eV_L\!\equiv\!\varepsilon\!=\!\mu$. 
Evidently, by this particular choice we exactly hit the Dirac point of the hole dispersion in $L$ and the electron dispersion in $R$. 
Since these states have zero momentum, $\bsy k\!=\!0$, they do not contribute to transport, completely suppressing LAR and CO. 

{\it Superconducting pairing.---}Next, we study the symmetry of the superconducting pairing potential that is proximity-induced into the TI-SC-TI device. To do so, we analytically compute the retarded Green function using a scattering state approach, see \cite{SuppMat}. In the basis defined above, this Green function can be written as $\bsy{\mathcal{G}}^r(x,x',y,y',k_y,\omega)\!=\!\bsy{\mathcal{G}}^r(x,x',k_y,\omega)\E{ik_y(y-y')}$ with
\begin{align}
\label{eq:4}
\bsy{\mathcal G}^r(x,x',k_y,\omega)&=
\begin{pmatrix}
\bsy G_{ee}^r & \bsy G_{eh}^r \\ \bsy G_{he}^r & \bsy G_{hh}^r
\end{pmatrix}, \\
\bsy G_{nm}^r(x,x',k_y,\omega)&=
\begin{pmatrix}
G_{nm}^{r,\uparrow\uparrow} & G_{nm}^{r,\uparrow\downarrow} \\ G_{nm}^{r,\downarrow\uparrow} & G_{nm}^{r,\downarrow\downarrow}
\end{pmatrix}.
\end{align}
Dependencies inside the matrices are omitted for ease of notation. 
Here, $\omega\!=\!\varepsilon\!+\!i0^+$ is the energy shifted infinitesimally into the positive complex plane to impose outgoing boundary conditions \cite{Datta07}. 

For convenience, we define the anomalous Green function in a rotated basis as
\begin{align}
\bsy{\mathcal{F}}^r(x,x',k_y,\omega)\equiv-i \bsy G_{eh}^r \bsy \sigma_y=
\begin{pmatrix}
G_{eh}^{r,\uparrow\downarrow} & -G_{eh}^{r,\uparrow\uparrow} \\ G_{eh}^{r,\downarrow\downarrow}&  -G_{eh}^{r,\downarrow\uparrow}
\end{pmatrix} ,
\end{align}
which we further decompose into singlet/triplet parts
\begin{align}
\label{eq:5}
\bsy{\mathcal{F}}^r(x,x',k_y,\omega)=\sum\limits_{i\in\{0,x,y,z\}}f_i^r(x,x',k_y,\omega)\bsy\sigma_i.
\end{align}
In this equation, $f_0^r$ is the singlet, and $f_z^r$ as well as $f^r_{\uparrow \uparrow / \downarrow \downarrow}\!\equiv\!\mp f^r_x\!+\!if^r_y$ are the triplet amplitudes.

Before we discuss specific results on the pairing amplitude for our setup, we state more general arguments based on symmetry. It is straightforward to show that $f_y^r$ and $f_z^r$ are odd in $k_y$, independently of the choice of $x$ and $x'$. Since we later on average all quantities with respect to $k_y$, they do not contribute to the (averaged) triplet pairing amplitudes \cite{Burset2015}. This implies that $f^r_{\uparrow \uparrow }$ and $f^r_{\downarrow \downarrow}$ only differ in their sign. The remaining amplitudes, $f_0^r$ and $f_x^r$, are even functions of $k_y$. Hence, they can (in principle) remain finite after averaging.

\begin{figure} 
\renewcommand{\figurename}{Fig.}
\centering
\includegraphics[width=.95\linewidth]{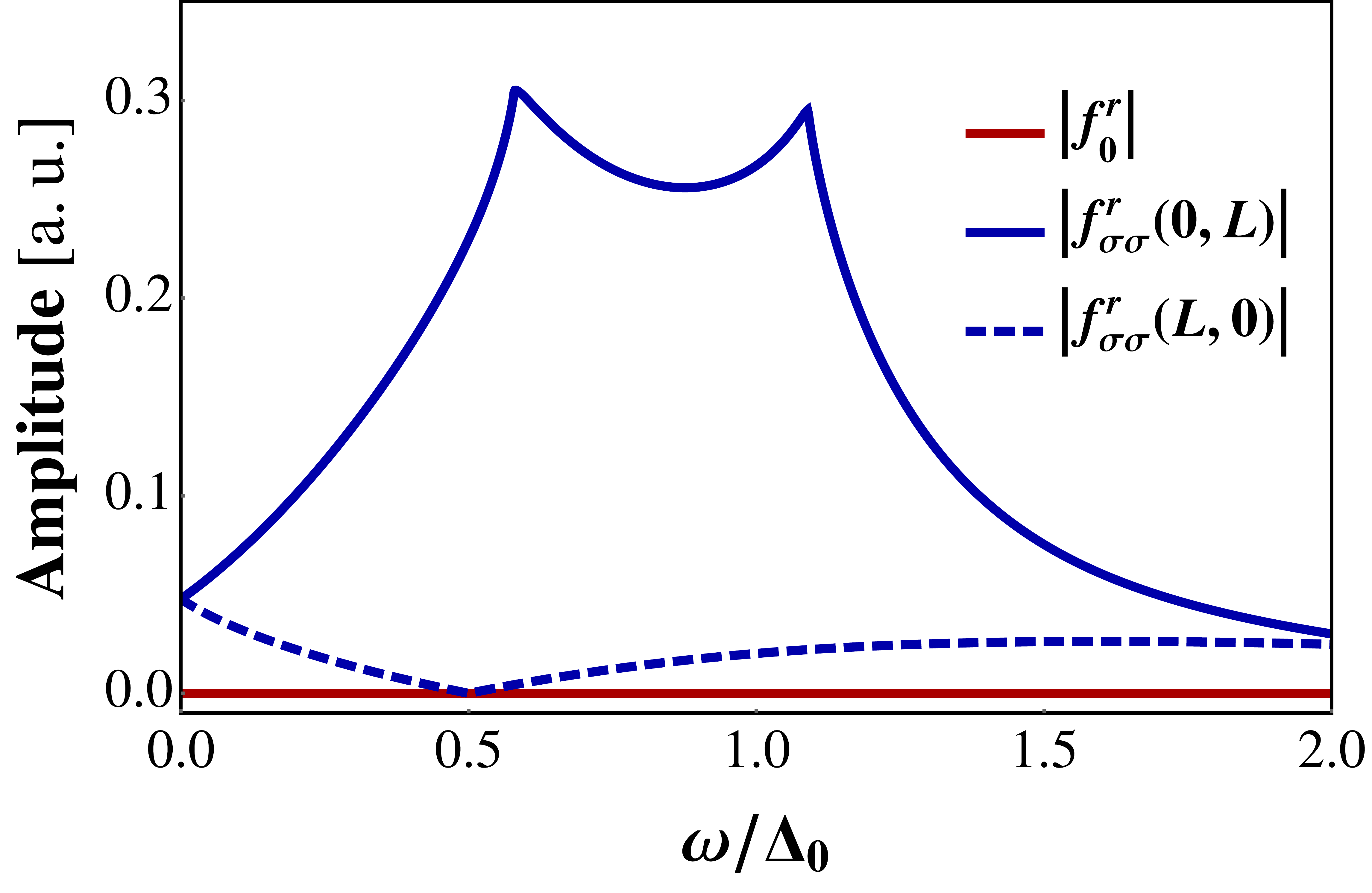}
\caption{ Averaged moduli of the nonlocal pairing amplitudes from one interface to the other one. For the bipolar setup, nonlocal singlet pairing is completely suppressed, while triplet pairing remains finite. We set $\mu\!=\!0.5\Delta_0$, $\mu_S\!=\!10\Delta_0$, and $L_S\!=\!1.1\xi$ ($\xi\!=\!v_F/\Delta_0$) with $\sigma\!\in\!\{\uparrow,\downarrow\}$. $f^r_0$ illustrates both $f_0^r(0,L)$ and $f_0^r(L,0)$.}
\label{fig:3}
\end{figure}

One of our main results is that this bipolar TI-SC-TI-junction acts as a perfect filter for nonlocal triplet pairing in the superconductor because of the helicity of the TI surface states. 
To substantiate this claim, we first present our results on the nonlocal singlet pairing amplitude. Labeling as $x_L\!\equiv\!x\!<\!0$ ($x_R\!\equiv\!x\!>\!L_S$) any point in region $L$ ($R$), we find \footnotemark[2]
\footnotetext[2]{The expressions in Eqs. \eqref{eq:6}-\eqref{eq:8} are derived under the condition $0<-\mu_R<\mu_L<\varepsilon$. They capture the spin amplitudes correctly as we approach the bipolar regime.}
\begin{align}
\label{eq:6}
f_0^r(x_L,x_R)=\sin\left(\frac{\theta_L^e-\theta_R^h}{2}\right) g_{eh}^{\uparrow\downarrow}(x_L,x_R), 
\end{align}
with
\begin{align}
g_{eh}^{\uparrow\downarrow}(x_L,x_R) =
\frac{\exp\left(i\frac {\theta_L^e+\theta_R^h}{2} \right)}{\cos\left(\theta^e_L\right)} \E{-i\left(k_L^ex_L-k_R^hx_R\right)}d_1.
\end{align}
For the nonlocal equal-spin triplet amplitude we get
\begin{align}
\label{eq:8}
f_{\uparrow\uparrow}^r(x_L,x_R)&=-\frac i2 \frac{\E{i\left(\theta_L^e+\theta_R^h\right)}}{\cos\left(\theta^e_L\right)}\E{i\left(k_R^hx_R-k_L^ex_L\right)}d_1.
\end{align}
In these equations, $d_1$ denotes the scattering amplitude related to an electron incident from $L$ being transmitted into $R$ as a hole, see \cite{SuppMat}. Similar equations to Eqs. \eqref{eq:6}-\eqref{eq:8} also apply to the choice $x_L \leftrightarrow x_R$. In the bipolar regime, $\mu_L=-\mu_R\equiv\mu$, we immediately deduce from the definition of the angles of incidence that $\theta^{e(h)}_L=\theta^{h(e)}_R$. 
As a consequence, the nonlocal spin singlet amplitude in Eq. \eqref{eq:6} evaluates to zero, making this pairing absent across the junction independently of frequency $\omega$ and mode index $k_y$ \footnotemark[1]. 
In contrast, the nonlocal spin triplet amplitude in Eq. \eqref{eq:8} remains finite, see Fig. \ref{fig:3}. 
\footnotetext[1]{The local spin singlet, where the two electrons forming the Cooper pair are taken from the same region, is proportional to the LAR amplitude. It is thus also suppressed at the bipolar junction. }

\begin{figure} 
\renewcommand{\figurename}{Fig.}
\centering
\includegraphics[width=.95\linewidth]{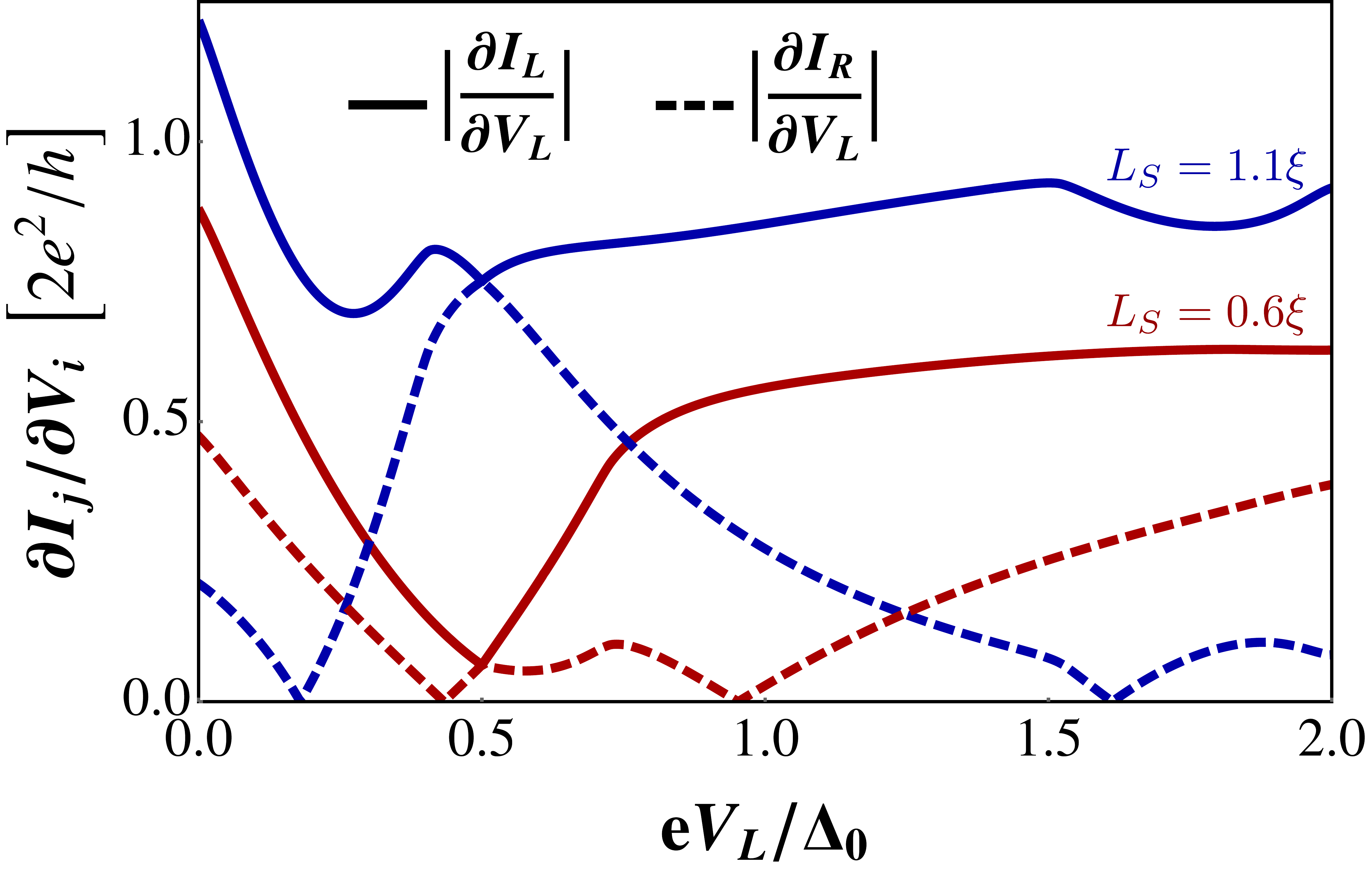}
\caption{Averaged moduli of the local and nonlocal conductance
 for $L_S\!=\!1.1\xi$ (blue lines) and $L_S\!=\!0.6\xi$ (red lines). 
 At $\mathrm{e} V_L\!=\!\mu$, the curves touch, which is a characteristic feature of the bipolar setup. We set $\mu\!=\!0.5\Delta_0$ and $\mu_S\!=\!10\Delta_0$.}
 \label{fig:4}
\end{figure}

{\it Transport properties.---}After the identification of pure triplet pairing across the junction, we want to use this particular property for an application in superconducting spintronics. To achieve this goal, we need to look at non-equilibrium transport properties of the TI-SC-TI junction by means of the (extended) Blonder-Tinkham-Klapwijk (BTK) formalism \cite{Blonder1982a,Anantram_1996,Lambert_1998,Falci2001,Linder2008}. This allows for the determination of the probability amplitudes of the scattering processes, in general, and the local and nonlocal conductance, in particular. A detailed evaluation of the probability amplitudes, see \cite{SuppMat}, confirms the physical picture we elaborated in Fig. \ref{fig:2}. For an electron excited in $L$,  both the amplitudes for LAR ($\equiv\! R_1^{eh}$) and CO ($\equiv\! T_1^{ee}$) vanish at $eV_L\!=\!\mu$, while those for NR ($\equiv\! R_1^{ee}$) and CAR ($\equiv\! T_1^{eh}$) remain finite in the bipolar setup.
This property has a striking effect on the non-linear conductance $\partial I_j/\partial V_i$ at zero temperature $T\!=\!0$. Changing only $V_L$ and setting $V_R\!=\!0$, the non-linear conductance reduces to distinct local and nonlocal parts,
\begin{align}
\frac{\partial I_L}{\partial V_L}&=\frac{2e^2}{h}\left[ 1-R_1^{ee}(\mathrm{e} V_L,k_y)+R_1^{eh}(\mathrm{e} V_L,k_y)\right], \\
\frac{\partial I_R}{\partial V_L}&=\frac{2e^2}{h}\left[ T_1^{ee}(\mathrm{e} V_L,k_y)-T_1^{eh}(\mathrm{e} V_L,k_y)\right].
\end{align}
The unitarity of the scattering matrix, i.e. $R_1^{ee}\!+\!R_1^{eh}\!+\!T_1^{ee}\!+\!T_1^{eh}\!=\!1$, 
yields $T_1^{eh}(\mathrm{e} V_L\!=\!\mu,k_y)\!=\!1\!-\!R_1^{ee}(\mathrm{e} V_L\!=\!\mu,k_y)$ and thus
\begin{align}
\frac{\partial I_L}{\partial V_L}(\mathrm{e} V_L=\mu,k_y)&=-\frac{\partial I_R}{\partial V_L}(\mathrm{e} V_L=\mu,k_y).
\end{align}
At this particular point, corresponding to the choice of bias $\mathrm{e} V_L\!=\!\mu$ in our setup, the local and the nonlocal conductance coincide in their absolute value, see Fig. \ref{fig:4}. 
This property is very unusual in superconducting devices because, in general, the local conductance dominates its nonlocal counterpart. 
Thus, we do not only observe a particular behavior in the nonlocal superconducting pairing, but we are also able to pinpoint a sweet spot at $\mathrm{e} V_L\!=\!\mu$ with striking features in the transport that we can connect to the pumped spin into the SC.

\begin{figure} 
\renewcommand{\figurename}{Fig.}
\centering
\includegraphics[width=.95\linewidth]{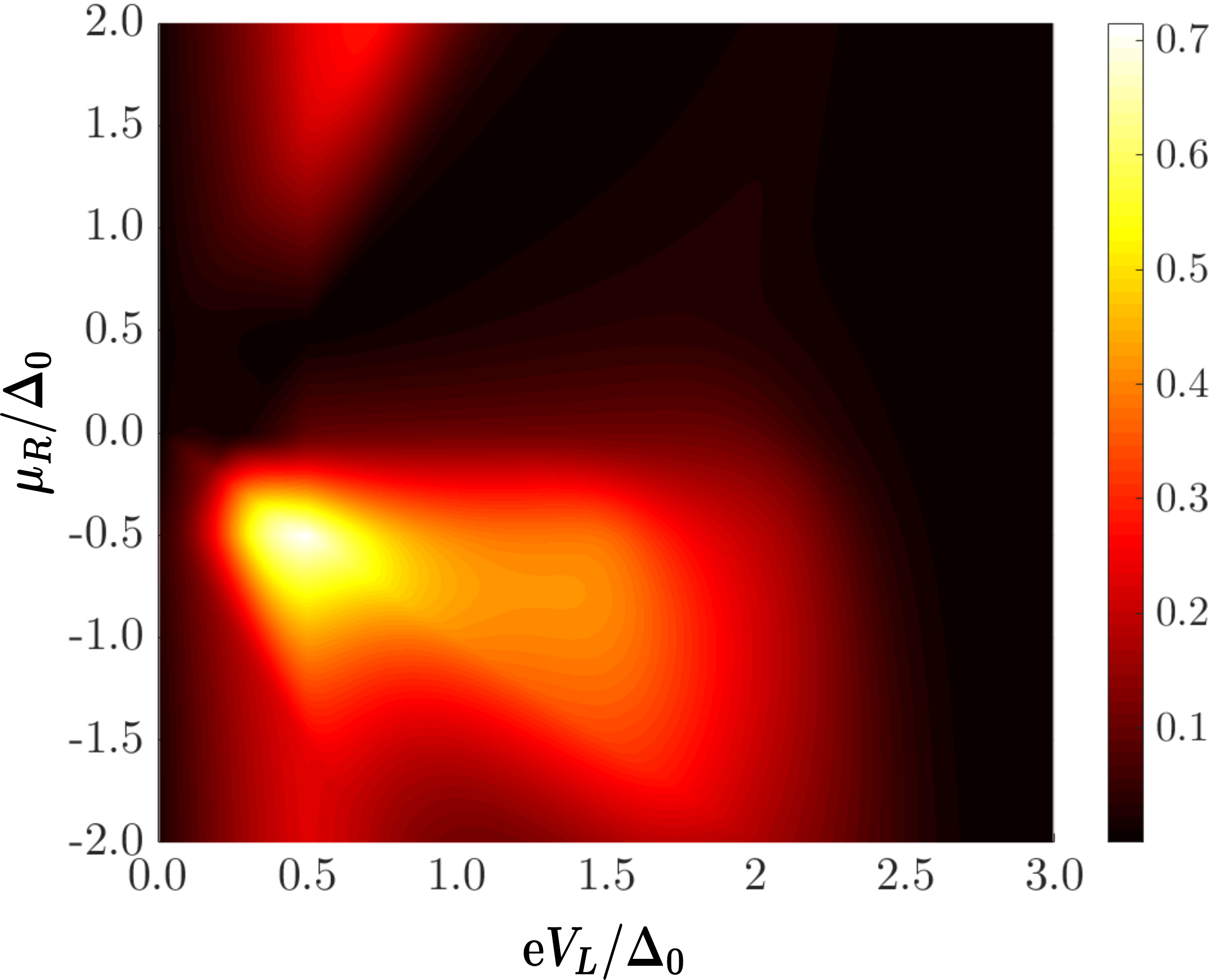}
\caption{Non-equilibrium net spin $\bsy{ \mathcal{S}}_x$ pumped into the SC as a function of $\mu_R$ and $V_L$. We set $\mu\!=\!0.5\Delta_0$, $\mu_S\!=\!10\Delta_0$, $L_S\!=\!1.1\xi$. }
\label{fig:5}
\end{figure}

{\it Spin injection.---}To do so, we quantify the non-equilibrium spin pumped into the SC via Andreev reflection processes introducing the quantity $\bsy{ \mathcal{S}}$, which we denote the non-equilibrium net spin pumped into the SC,
\begin{align}
\label{eq:10}
\bsy{ \mathcal{S}}=R_{LAR}\braket{\hat S}_{l}+T_{CAR}\braket{\hat S}_{nl},
\end{align}
with the local $\braket{ \cdot}_l$ and nonlocal $\braket{ \cdot}_{nl}$ contributions
\begin{align}
\label{eq:11}
\braket{\hat S}_{l}=\braket{\hat S}+\braket{\hat S}_{LAR}, &&
\braket{\hat S}_{nl}=\braket{\hat S}+\braket{\hat S}_{CAR}.
\end{align}
Here, $\hat S$ is the Cooper pair spin operator (see \cite{SuppMat})
and $\braket{\hat S}$, $\braket{\hat S}_{LAR}$ and $\braket{\hat S}_{CAR}$ are the spin expectation values of the incident, the Andreev reflected, and the crossed Andreev reflected particle, respectively.  Let us explain why this quantity is a good measure for the spin pumped into the system through triplet Cooper pairs. If we consider, for instance, an electron incident from $L$ moving towards the interface at $x\!=\!0$, then there are two processes resulting in the emergence of non-equilibrium Cooper pairs in the SC, 
\begin{inparaenum}[(i)]
	\item LAR with probability $R_{LAR}(=\!R_1^{eh})$, where the spins of the incident electron and the reflected hole, i.e. $\braket{\hat S}+\braket{\hat S}_{LAR}$, are transferred to the SC; and
	\item CAR with probability $T_{CAR}(=\!T_1^{eh})$, where the spin of the transmitted hole is added to that of the electron, i.e. $\braket{\hat S}+\braket{\hat S}_{CAR}$. 
\end{inparaenum}
Thus, a good estimate of the net spin of the Cooper pairs pumped into the SC via Andreev reflection is the sum of both contributions, $\braket{\hat S}_{l}$ and $\braket{\hat S}_{nl}$, weighted with their respective probability amplitudes.
Explicitly, we obtain (under the choice $\hbar\!=\!1$)
\begin{subequations}\label{eq:12}
\begin{align}
\braket{\hat S}=& \frac{\sign(\varepsilon+\mu_L)}{2}\Big(\zeta_L^e \cos \theta_L^e,\sin \theta_L^e,0\Big)^T, \\
\label{eq:13}
\braket{\hat S}_{LAR}=& \frac{\sign(\varepsilon-\mu_L)}{2}\Big(-\zeta_L^h \cos \theta_L^h,\sin \theta_L^h,0\Big)^T, \\
\label{eq:14}
\braket{\hat S}_{CAR}=&\frac{\sign(\varepsilon-\mu_R)}{2}\Big(\zeta_R^h \cos \theta_R^h,\sin \theta_R^h,0\Big)^T.
\end{align}
\end{subequations}
The norm of these quantities is always $1/2$, as expected for fermions, and their $z$-component vanishes. Moreover, their $y$-component is odd under $\theta_i^{e/h}$, such that its average with respect to $k_y$ will vanish. Thus, only the $x$-component of the net spin (pumped in the SC after averaging over all angles of incidence) remains finite. 
We can therefore focus on this part of Eqs. \eqref{eq:12}. Indeed, we find a bound for this component of the net spin
\begin{align}
0\le\abs{\bsy{ \mathcal{S}}_x}\le R_{LAR}+T_{CAR}\le 1,
\end{align}
which corresponds to the maximal angular momentum transferred per scattering event.
 If we plot this quantity as a function of both $\mu_R$ and $\mathrm e V_L$, see Fig. \ref{fig:5}, we find that, in the bipolar setup and for our choice of the SC length $L_S$, we pump the largest amount of non-equilibrium net spin into the SC by tuning the bias in the vicinity of the sweet spot $\mathrm{e} V_L\!=\!\mu$. 

{\it Summary.---}The breaking of spin rotational invariance in materials with strong spin-orbit locking reveals striking physics. We found that, by connecting a $p$-type to an $n$-type TI via a SC domain, this bipolar TI-SC-TI junction acts as an effective nonlocal spin triplet filter, where nonlocal singlet pairing is completely suppressed. Due to the helicity of the TI states, the non-equilibrium triplet Cooper pairs pumped into the SC carry a finite net spin. We thus propose this setup as an all-electric nanostructure for applications in spintronics. The effect of the spin pumping is strongest in the vicinity of the sweet spot $\mathrm eV_L\!=\!\mu_L\!=\!-\mu_R$, where we estimate a net spin of, depending on the bias range, approximately $\bsy{ \mathcal{S}}_x\!\approx\! 0.6$--$1.2\hbar$ per unit volume (see \cite{SuppMat}) for a SC length of $L_S\!=\!1.1\xi$. This sweet spot is experimentally easy to detect, since the local and the nonlocal conductivities for a bias applied to the left TI coincide in their moduli. 
Interestingly, pumped equal-spin Cooper pairs out of equilibrium into a superconductor could lead to long-range spin accumulation \cite{Heikkila_2015,Tero_review}. 
Connecting a second superconductor to the central region of the bipolar junction, the spin accumulation could be measured through local SQUID or Hall probes \cite{Maeno_2005,Maeno_2007,Yakovenko_2008}.
\\
We thank Y. Asano, F. S. Bergeret, F. Crepin, F. Dominguez, J. Linder, J. Pekola, B. Scharf, S. Zhang and N. Traverso Ziani for interesting discussions. 
Financial support by the DFG (SPP1666 and SFB1170 "ToCoTronics"), the Helmholtz Foundation (VITI), and the ENB Graduate School on "Topological Insulators" is gratefully acknowledged. 
P.B. acknowledges funding from the European Union's Horizon 2020 research and innovation programme under the Marie Sk\l odowska-Curie grant agreement No. 743884. 
\bibliographystyle{apsrev4-1}

\appendix
\onecolumngrid
\newpage
\renewcommand{\theequation}{A.\arabic{equation}}
\addto\captionsenglish{\renewcommand{\figurename}{\textbf{Fig. A}}}
\setcounter{equation}{0}
\section*{Supplemental material}

In this supplementary material, we introduce the definitions for the eigenstates and scattering states used to derive all the quantities in the main text. The methods to obtain the Green function as well as the transport properties are illustrated and some significant results are presented explicitly. 

\section{Preliminary calculations and definitions}
\noindent Starting from Schrödinger's equation
\begin{align}
\label{eq:A1}
\Hamil_{\mathrm{BdG}} \bsy\Psi(x,y)=\varepsilon\, \bsy\Psi(x,y)
\end{align}
with the Bogoliubov--de Gennes (BdG) and Bloch Hamiltonian introduced in Eqs. (1) and (2) of the main text,
\begin{align}
\label{eq:A2}
\Hamil_{\mathrm{BdG}}=
\begin{pmatrix}
H_0(\bsy k) &&i\Delta(x) \p y \\
-i\Delta(x) \p y && -H_0^*(-\bsy k)
\end{pmatrix}, && 
H_0(\bsy k)&=v_F\left(\hat k_x\,\p x+\hat k_y \,\p y\right)-\mu(x)\,\p 0, && \bsy k=\left(\hat k_x,\hat k_y\right)^T,
\end{align}
and the definitions of the electrochemical and superconducting (SC) pairing potential,
\begin{align}
\mu(x)=\mu_L\, \Theta(-x)+\mu_S\,\Theta(x)\Theta(L_S-x)+\mu_R\,\Theta(x-L_S), &&
\Delta(x)=\Delta_0\,\Theta(x)\Theta(L_S-x),
\end{align}
we make use of the translational invariance along the $y$--axis and perform a partial Fourier transformation in this variable,
\begin{align}
\bsy\Psi(x,y)=\int\limits_{-\infty}^\infty \bsy\psi(x,k_y)\E{ik_yy}\D k_y.
\end{align}
As a result, $H_0(\bsy k)$ reduces to a quasi--1D operator, which simplifies further calculations. From here, we obtain the eigenstates of each domain by solving Eq. (1) in its quasi--1D representation. This yields right and left moving states 
\begin{align}
\label{eq:A3}
\bsy{\psi}_{\stackrel{\rightarrow}{e/h}}^L(x)&=\frac{1}{\sqrt{2}}\left(\frac{\bar{ k}_L^{e/h}\mp ik_y}{\varepsilon\pm\mu_L}\bsy{\hat e}_{1/3}+\bsy{\hat e}_{2/4}\right)^T\E{i\bar{ k}_L^{e/h}x}, &&
\bsy{\psi}_{\stackrel{\leftarrow}{e/h}}^L(x)=\frac{1}{\sqrt{2}}\left(\frac{k_L^{e/h}\pm ik_y}{\varepsilon\pm\mu_L}\bsy{\hat e}_{1/3}-\bsy{\hat e}_{2/4}\right)^T\E{-ik_L^{e/h}x}, 
\end{align}
in the left ($L$) topological insulator (TI) and, correspondingly,
\begin{align}
\label{eq:A4}
\bsy{\psi}_{\stackrel{\rightarrow}{e/h}}^R(x)&=\frac{1}{\sqrt{2}}\left(\frac{k_R^{e/h}\mp ik_y}{\varepsilon\pm\mu_R}\bsy{\hat e}_{1/3}+\bsy{\hat e}_{2/4}\right)^T\E{ik_R^{e/h}x},&&
\bsy{\psi}_{\stackrel{\leftarrow}{e/h}}^R(x)=\frac{1}{\sqrt{2}}\left(\frac{\bar{ k}_R^{e/h}\pm ik_y}{\varepsilon\pm\mu_R}\bsy{\hat e}_{1/3}-\bsy{\hat e}_{2/4}\right)^T\E{-i\bar{ k}_R^{e/h}x}, 
\end{align}
in the right ($R$) TI. In the superconductor (SC), we find
\begin{align}
\label{eq:A5}
\bsy \psi^S_{1/2}(x)&=\mathcal{N}_{eq/hq}
\begin{pmatrix}
(\varepsilon\pm\Omega)(k^{eq/hq}-ik_y) \\
(\varepsilon\pm\Omega)(\mu_S\pm\Omega)\\
-(\mu_S\pm\Omega)\Delta_0 \\
(k^{eq/hq}-ik_y) \Delta_0
\end{pmatrix} 
\E{ik^{eq/hq}x} , &&
\bsy \psi^S_{3/4}(x)=\mathcal{N}_{eq/hq}
\begin{pmatrix}
(\varepsilon\pm\Omega)(k^{eq/hq}+ik_y) \\
-(\varepsilon\pm\Omega)(\mu_S\pm\Omega)\\
(\mu_S\pm\Omega)\Delta_0 \\
(k^{eq/hq}+ik_y) \Delta_0
\end{pmatrix} 
\E{-ik^{eq/hq}x}
\end{align}
with the normalization factors
\begin{align}
\mathcal{N}_{eq}^{-1}=2\sqrt{\varepsilon(\varepsilon+\Omega)}(\mu_S+\Omega), && \mathcal{N}_{hq}^{-1}=2\sqrt{\varepsilon(\varepsilon-\Omega)}(\mu_S-\Omega).
\end{align}
To obtain a complete solution, each of these states needs to be multiplied by the plane wave $\chi(y)=\E{ik_yy}$. Here, the alphabetical indices distinguish electrons ($e$) from holes ($h$) in the TIs and electron--like ($eq$) from hole--like ($hq$) quasi--particles in the SC, while the arrows imply the direction of motion with respect to the $x$--axis. $\bar{(\cdot)}$ is the complex conjugate of the quantity $(\cdot)$ and $\{\bsy{\hat e}_l\}_{l \in \{1,2,3,4\}}$ is the complete, orthonormal set of the standard basis vectors of $\mathbb{R}^4$.  The wave vectors in the TIs and the SC read 
\begin{align}
\label{eq:A5a}
k^{e/h}_{i}=\zeta^{e,h}_i\sqrt{(\varepsilon\pm\mu_i)^2-k_y^2}, &&
k^{eq/hq}=\zeta^{eq/hq}\sqrt{\left(\mu_S\pm\Omega\right)^2-k_y^2},
\end{align}
with
\begin{align}
\zeta^{e,h}_i=\sign\left(\varepsilon\pm\mu_i+\abs{k_y}\right), &&\zeta^{eq/hq}=\sign\left(\varepsilon\pm\sqrt{\left(\mu_S+\abs{k_y}\right)^2+\Delta_0^2}\right), &&
\Omega=\begin{cases} \sign(\varepsilon)\sqrt{\varepsilon^2-\Delta_0^2}, &\abs \varepsilon\ge\Delta_0 \\ i \sqrt{\Delta_0^2-\varepsilon^2},&\abs\varepsilon<\Delta_0\end{cases}.
\end{align}
Note that, in the main text, we have used an angular representation for argumentative reasons. Definitions of the angles can be found there. We intentionally do not use such a representation here, since this renders the $k_y$--integration as well as the determination of the retarded Green function more apparent.

We now define the scattering states for the TI--SC--TI nanostructure introduced in the main text, i.e.,
\begin{align}
\bsy\phi_{1/2}(x)&=
\begin{cases}
\bsy{\psi}_{\stackrel{\rightarrow}{e/h}}^L(x)+a_{1/2}\bsy{\psi}_{\stackrel{\leftarrow}{h/e}}^L(x)+b_{1/2}\bsy{\psi}_{\stackrel{\leftarrow}{e/h}}^L(x),&x<0 \\
\sum\limits_{i=1}^4s_{1/2}^i\bsy \psi_i^s(x), & 0<x<L_S\\
c_{1/2}\bsy{\psi}_{\stackrel{\rightarrow}{e/h}}^R(x)+d_{1/2}\bsy{\psi}_{\stackrel{\rightarrow}{h/e}}^R(x),&x>L_S
\end{cases}, \nn \\ 
\label{eq:A6}
 \\
\bsy\phi_{3/4}(x)&=
\begin{cases}
c_{3/4}\bsy{\psi}_{\stackrel{\leftarrow}{e/h}}^L(x)+d_{3/4}\bsy{\psi}_{\stackrel{\leftarrow}{h/e}}^L(x),&x<0  \\
\sum\limits_{i=1}^4s_{3/4}^i\bsy \psi_i^s(x), & 0<x<L_S\\
\bsy{\psi}_{\stackrel{\leftarrow}{e/h}}^R(x)+a_{3/4}\bsy{\psi}_{\stackrel{\rightarrow}{h/e}}^R(x)+b_{3/4}\bsy{\psi}_{\stackrel{\rightarrow}{e/h}}^R(x),&x>L_S
\end{cases}. \nn
\end{align}
Here, $\bsy\phi_{1(2)}(x)$ describes the scattering processes based on an electron(hole) incident from the left, while $\bsy\phi_{3(4)}(x)$ describes the same for an electron(hole) incident from the right. $a$ is the amplitude for an incident particle being Andreev reflected at the interface (LAR), $b$ the one for normal reflection (NR). Electron or hole co--tunneling (CO) is linked to the amplitude $c$, while crossed Andreev reflection (CAR) goes with $d$. These scattering coefficients are determined by matching the wave functions at the interfaces,
\begin{align}
\bsy\phi_l(x=0-0^+)=\bsy\phi_l(x=0+0^+), && \bsy\phi_l(x=L_S-0^+)=\bsy\phi_l(x=L_S+0^+).
\end{align}
To calculate the retarded Green function, we need to find as well the eigenstates of the transposed \cite{McMillan1968, Tanaka2000} BdG Hamiltonian in Eq. \eqref{eq:A2}, which reads as
\begin{align}
\label{eq:A7}
\Hamil_{\mathrm{BdG}}^T=
\begin{pmatrix}
H_0^*(-\bsy k) &&i\Delta(x) \p y \\
-i\Delta(x) \p y && -H_0(\bsy k)
\end{pmatrix}.
\end{align}
In this context, we need to define the partial Fourier transform as
\begin{align}
\tilde{\bsy\Psi}(x,y)=\int\limits_{-\infty}^\infty \tilde{\bsy\psi}(x,k_y)\E{-ik_yy}\D k_y.
\end{align}
We denote $\tilde{\bsy\psi}(x,k_y)$ as the transposed eigenstates. Explicitly, they are given by
\begin{align}
\label{eq:A8}
\bsy{\tilde\psi}_{\stackrel{\rightarrow}{e/h}}^L(x)&=\frac{1}{\sqrt{2}}\left(\frac{\bar{ k}_L^{e/h}\mp ik_y}{\varepsilon\pm\mu_L}\bsy{\hat e}_{1/3}-\bsy{\hat e}_{2/4}\right)^T\E{i\bar{ k}_L^{e/h}x}, &&
\bsy{\tilde\psi}_{\stackrel{\leftarrow}{e/h}}^L(x)=\frac{1}{\sqrt{2}}\left(\frac{k_L^{e/h}\pm ik_y}{\varepsilon\pm\mu_L}\bsy{\hat e}_{1/3}+\bsy{\hat e}_{2/4}\right)^T\E{-ik_L^{e/h}x}, \\
\label{eq:A9}
\bsy{\tilde\psi}_{\stackrel{\rightarrow}{e/h}}^R(x)&=\frac{1}{\sqrt{2}}\left(\frac{k_R^{e/h}\mp ik_y}{\varepsilon\pm\mu_R}\bsy{\hat e}_{1/3}-\bsy{\hat e}_{2/4}\right)^T\E{ik_R^{e/h}x},&&
\bsy{\tilde\psi}_{\stackrel{\leftarrow}{e/h}}^R(x)=\frac{1}{\sqrt{2}}\left(\frac{\bar{ k}_R^{e/h}\pm ik_y}{\varepsilon\pm\mu_R}\bsy{\hat e}_{1/3}+\bsy{\hat e}_{2/4}\right)^T\E{-i\bar{ k}_R^{e/h}x}, 
\end{align}
in $L$ and $R$, respectively, and
\begin{align}
\label{eq:A10}
\bsy{ \tilde \psi}^S_{1/2}(x)&=\mathcal{N}_{eq/hq}
\begin{pmatrix}
(\varepsilon\pm\Omega)(k^{eq/hq}-ik_y) \\
-(\varepsilon\pm\Omega)(\mu_S\pm\Omega)\\
(\mu_S\pm\Omega)\Delta_0 \\
(k^{eq/hq}-ik_y) \Delta_0
\end{pmatrix} 
\E{ik^{eq/hq}x} , &&
\bsy{\tilde \psi}^S_{3/4}(x)=\mathcal{N}_{eq/hq}
\begin{pmatrix}
-(\varepsilon\pm\Omega)(k^{eq/hq}+ik_y) \\
(\varepsilon\pm\Omega)(\mu_S\pm\Omega)\\
(\mu_S\pm\Omega)\Delta_0 \\
(k^{eq/hq}+ik_y) \Delta_0
\end{pmatrix} 
\E{-ik^{eq/hq}x},  
\end{align}
in the SC. With this, we define the transposed scattering states $\tilde{\bsy \phi_l}$ by replacing $\bsy\psi$ by $\tilde{\bsy\psi}$ and $a_l,b_l,c_l,d_l,s^i_l$ by $\tilde a_l,\tilde b_l,\tilde c_l,\tilde d_l,\tilde s^i_l$ in Eq. \eqref{eq:A6}, with $l\in\{1,2,3,4\}$. Note that these scattering coefficients can easily be related to each other by means of the Wronskian determinant \cite{Datta07} and Liouville's formula so to obtain
\begin{align}
\tilde{a}_l=-a_l, && \tilde{b}_l=b_l, && \tilde{c}_l=c_l,&& \tilde{d}_l=-d_l, && \tilde{s}_l^i=(-1)^ls_l^i.
\end{align}
For a given excitation energy $\varepsilon$, we average an arbitrary quantity $q(\varepsilon,k_y)$ by means of the integral
\begin{align}
\label{eq:A10a}
\braket{q(\varepsilon,k_y)}_{k_y}\equiv\frac{1}{2 k_{in}}\int\limits_{-k_{in}'}^{k_{in}'}q(\varepsilon,k_y)\,\D k_y.
\end{align}
Here, $k_{in}$ is the modulus of the wave vector of the incident particle at energy $\varepsilon$ and $k_y=0$, i.e., $k_{in}=k^{e/h}_i\Big|_{k_y=0}$. Intuitively, the limits of the integral should be $-k_{in}$ and $k_{in}$, however, one needs to take the Fermi wave vector mismatch (\textit{cf.} \cite{Kashiwaya1996,Linder2008}) into account, which eventually reduces the latter values to a critical wave vector (corresponding to a critical angle \cite{Cayssol2008}) $k_{in}'\le k_{in}$. 
\section{Green function}
\noindent We are now able to express the retarded Green function by means of outer products of the scattering states,
\begin{align}
\label{eq:A11}
\mathcal{G}^R(x,x')=
\begin{cases}
\alpha_1\bsy\phi_3(x)\tilde{\bsy\phi}_1^T(x')+\alpha_2\bsy\phi_3(x)\tilde{\bsy\phi}_2^T(x') +
\alpha_3\bsy\phi_3(x)\tilde{\bsy\phi}_1^T(x')+\alpha_4\bsy\phi_3(x)\tilde{\bsy\phi}_2^T(x'),&x<x'
\\ \\
\beta_1\bsy\phi_1(x)\tilde{\bsy\phi}_3^T(x')+\beta_2\bsy\phi_1(x)\tilde{\bsy\phi}_4^T(x') +
\beta_3\bsy\phi_2(x)\tilde{\bsy\phi}_3^T(x')+\beta_4\bsy\phi_2(x)\tilde{\bsy\phi}_4^T(x'),&x>x'
\end{cases}.
\end{align}
The coefficients $\alpha_l$ and $\beta_l$ are determined by requiring the discontinuity of $\mathcal{G}^R(x,x')$ at $x=x'$,
\begin{align}
\label{eq:A12}
\mathcal{G}^R(x,x')\Big|_{x=x'+0^+}-\mathcal{G}^R(x,x')\Big|_{x=x'-0^+}&=-i \left(\bsy \tau_0\otimes\bsy\sigma_x\right)
\end{align}
and read
\begin{align}
\alpha_1&=\beta_1=-i\,\frac{\omega+\mu_L}{\mathrm{Re}(k_L^e)}\frac{c_4}{c_3c_4-d_3d_4}=-i\,\frac{\omega+\mu_R}{\mathrm{Re}(k_R^e)}\frac{c_2}{c_1c_2-d_1d_2},\\
\alpha_2&=-\beta_2=i\,\frac{\omega-\mu_L}{\mathrm{Re}(k_L^h)}\frac{d_4}{c_3c_4-d_3d_4}=-i\,\frac{\omega+\mu_R}{\mathrm{Re}(k_R^e)}\frac{d_1}{c_1c_2-d_1d_2}, \\
\alpha_3&=-\beta_3=i\,\frac{\omega+\mu_L}{\mathrm{Re}(k_L^e)}\frac{d_3}{c_3c_4-d_3d_4}=-i\,\frac{\omega-\mu_R}{\mathrm{Re}(k_R^h)}\frac{d_2}{c_1c_2-d_1d_2},\\
\alpha_4&=\beta_4=-i\,\frac{\omega-\mu_L}{\mathrm{Re}(k_L^h)}\frac{c_3}{c_3c_4-d_3d_4}=-i\,\frac{\omega-\mu_R}{\mathrm{Re}(k_R^h)}\frac{c_1}{c_1c_2-d_1d_2}.
\end{align}
The first expression is obtained by solving Eq. \eqref{eq:A12} in the domain $x<0$, the second in the domain $x>L_S$. The coefficients $\alpha_l$ and $\beta_l$ take the same value along the junction. However, it is advantageous to know their explicit analytical expressions in terms of the parameters of each domain. With this, we obtain the nonlocal, retarded anomalous Green functions as
\begin{align}
\mathcal{G}^R_{eh}(x<0,x'>L_S)&=\frac{i\,\E{-i\left(k_L^ex-k_R^hx'\right)}d_1}{2\mathrm{Re}(k_L^e)(\omega-\mu_R)}
\begin{pmatrix}
(k_L^e+i\,k_y)(k_R^h+i\,k_y) && -(k_L^e+i\,k_y)(\omega-\mu_R) \\ \\
-(k_R^h+i\,k_y)(\omega+\mu_L) && (\omega+\mu_L)(\omega-\mu_R)
\end{pmatrix}, \\ \nn \\
\mathcal{G}^R_{eh}(x>L_S,x'<0)&=\frac{i\,\E{i\left(k_R^ex-k_L^hx'\right)}d_3}{2\mathrm{Re}(k_R^e)(\omega-\mu_L)}
\begin{pmatrix}
(k_L^h-i\,k_y)(k_R^e-i\,k_y) && (k_R^e-i\,k_y)(\omega-\mu_L) \\ \\
(k_R^e-i\,k_y)(\omega+\mu_R) && (\omega+\mu_R)(\omega-\mu_L)
\end{pmatrix}.
\end{align} 
In angular representation, we have
\begin{align}
\frac{k_i^{e/h}+ik_y}{\omega\pm\mu_i}=\zeta^{e/h}_i\sign(\omega\pm\mu_i)\E{i\zeta^{e/h}_i\theta^{e/h}_i}, &&
\frac{k_i^{e/h}-ik_y}{\omega\pm\mu_i}=\zeta^{e/h}_i\sign(\omega\pm\mu_i)\E{-i\zeta^{e/h}_i\theta^{e/h}_i},
\end{align}
 with $i\in\{L, R\}$ such that we obtain the pairing amplitudes in Eqs. (8)-(10) of the main text under the condition $0<-\mu_R<\mu_L<\varepsilon$.
\section{Transport properties}
\begin{figure}[htpb]
\renewcommand{\figurename}{Fig.}
\renewcommand\thefigure{A\arabic{figure}} 
\centering
\subfloat[{Bias ($V_L$) dependence.}]{\includegraphics[width=0.44\textwidth]{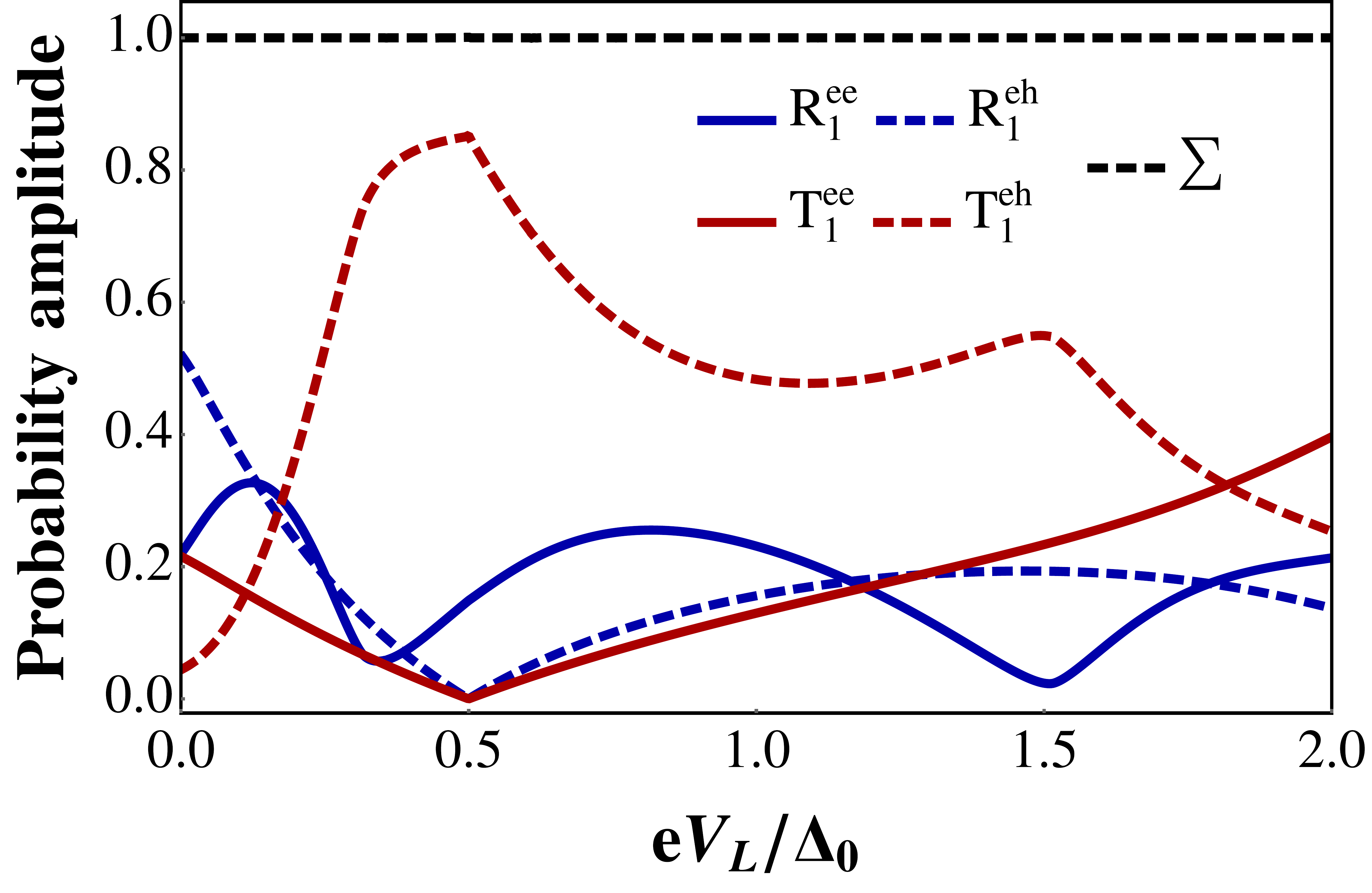}} \hfill
\subfloat[{SC length ($L_S$) dependence (only CAR).}]{\includegraphics[width=0.44\textwidth]{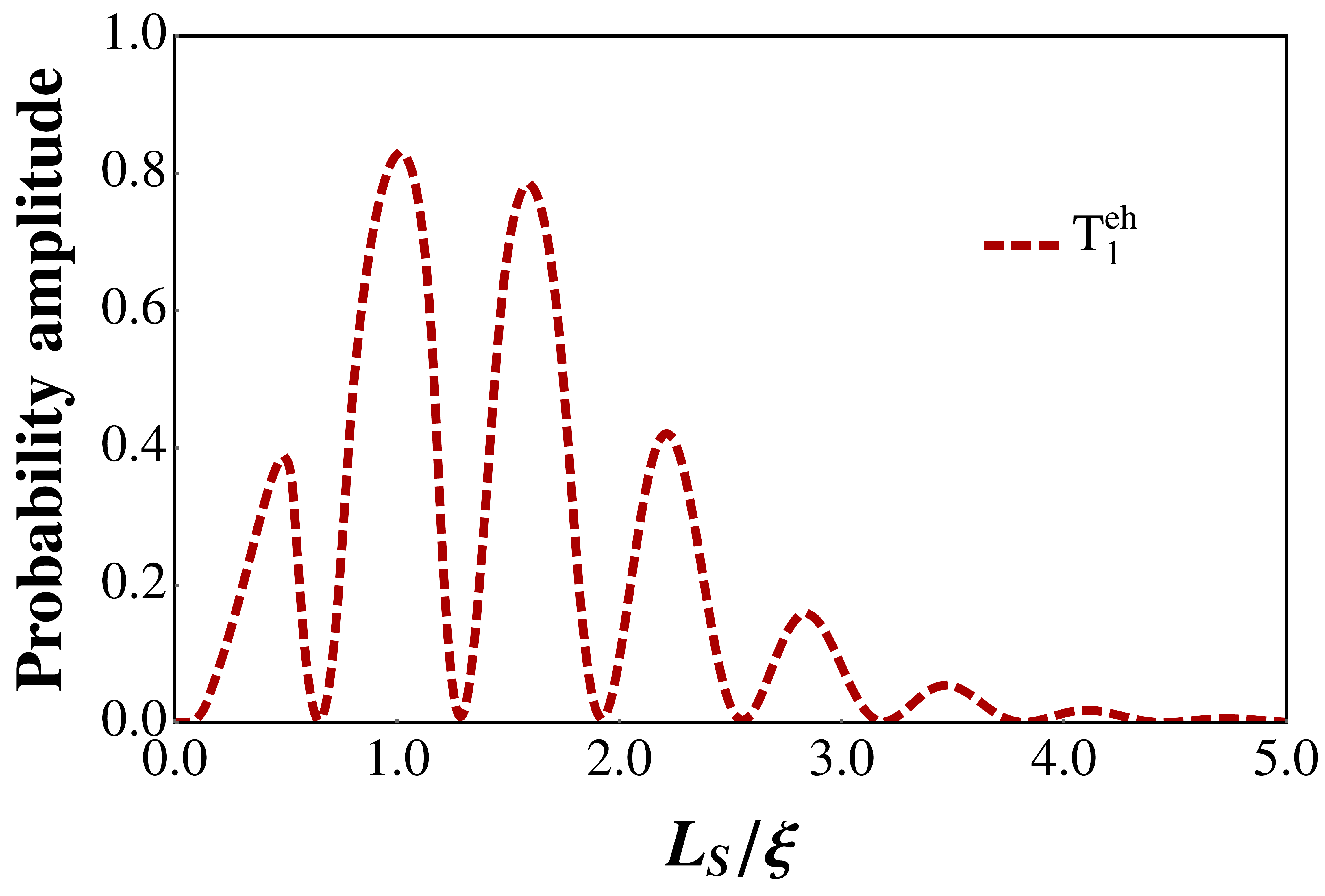}} 
\caption{Averaged transport probabilities for an electron excited in the left lead moving towards the SC interface as a function of \textbf{(a)} the bias applied to $L$ and \textbf{(b)} the length of the SC. Blue (red) lines indicate reflection (transmission) processes, solid (dashed) lines indicate processes preserving (changing) the particle type. The black, dashed line in (a) is the sum of the amplitudes $\bsy\Sigma$. We choose $\mu=0.5\Delta_0$, $\mu_S=10\Delta_0$, and \textbf{(a)} $L_S=1.1\xi$ \textbf{(b)} $\mathrm e V_L=\mu$ in the figure.}
\label{fig:A1}
\end{figure}

\noindent The probability amplitudes for each of the scattering processes are determined from the probability current density operator in the $x$--direction,
\begin{align}
\bsy j_x=\partial_{ \hat{k}_x}\Hamil_{\mathrm{BdG}}=\left(\bsy{\tau}_0\otimes\p x\right)
\end{align}
and read, normalized to the incident current, as
\begin{itemize}
\item electron excited in $L$ moving towards the interface at $x=0$ $\big(\hat = \;\bsy \phi_1(x)\big)$
\begin{align}
R^{ee}_1=\abs{b_1}^2, && R^{eh}_1=\frac{\varepsilon+\mu_L}{\varepsilon-\mu_L}\frac{\mathrm{Re}(k_L^h)}{k_L^e}\abs{a_1}^2, && T^{ee}_1=\frac{\varepsilon+\mu_L}{\varepsilon+\mu_R}\frac{\mathrm{Re}(k_R^e)}{k_L^e}\abs{c_1}^2, && T^{eh}_1=\frac{\varepsilon+\mu_L}{\varepsilon-\mu_R}\frac{\mathrm{Re}(k_R^h)}{k_L^e}\abs{d_1}^2.
\end{align}
\item electron excited in $R$ moving towards the interface at $x=L_S$ $\big(\hat = \;\bsy \phi_3(x)\big)$
\begin{align}
R^{ee}_3=\abs{b_3}^2, && R^{eh}_3=\frac{\varepsilon+\mu_R}{\varepsilon-\mu_R}\frac{\mathrm{Re}(k_R^h)}{k_R^e}\abs{a_3}^2, && T^{ee}_3=\frac{\varepsilon+\mu_R}{\varepsilon+\mu_L}\frac{\mathrm{Re}(k_L^e)}{k_R^e}\abs{c_3}^2, && T^{eh}_3=\frac{\varepsilon+\mu_R}{\varepsilon-\mu_L}\frac{\mathrm{Re}(k_L^h)}{k_R^e}\abs{d_3}^2.
\end{align}
\end{itemize}
As we assume to excite electrons for $\varepsilon>0$, we do not consider the amplitudes for excited holes here. With this, and by means of an extended Blonder-Tinkham-Klapwijk (BTK) formalism \cite{Blonder1982a, Anantram_1996, Lambert_1998, Falci2001,  Linder2008}, we obtain the conductance at $T=0$ as
\begin{align}
\frac{\partial I_{L/R}}{\partial V_{L/R}}=\frac{2e^2}{\hbar}\left[1-R^{ee}_{1/3}(\mathrm{e} V_L,k_y)+R^{eh}_{1/3}(\mathrm{e} V_L,k_y)\right], &&
\frac{\partial I_{R/L}}{\partial V_{L/R}}=\frac{2e^2}{\hbar}\left[T^{ee}_{1/3}(\mathrm{e} V_L,k_y)-T^{eh}_{1/3}(\mathrm{e} V_L,k_y)\right].
\end{align}
In Fig. \ref{fig:A1}, we plot the probability amplitudes for an electron excited in $L$ in the bipolar setup, averaged with respect 
\FloatBarrier
\noindent to $k_y$ according to Eq. \eqref{eq:A10a}. As explained in the main text, LAR and CO are completely suppressed at $\mathrm e V_L=\mu$, while NR and CAR remain finite, see Fig. \ref{fig:A1} \textbf{(a)}. This results in the particular behavior of the conductance in the sweet spot. Furthermore, the amplitudes oscillate as a function of the superconductor length, see Fig. \ref{fig:A1} \textbf{(b)}, where we plotted the CAR probability $T^{eh}_1(L_S)$. This stems from the fact that, in the subgap regime and for $\mu_S\gg\Delta_0$, the wave vectors $k^{eq/hq}$ in the SC, see Eq. \eqref{eq:A5a}, have finite real and imaginary parts such that we can observe quasi--particle interactions resulting in the oscillatory behavior of the transport probabilities.

\section{Spin injection}
\begin{figure} [ht]
\renewcommand{\figurename}{Fig.}
\renewcommand\thefigure{A\arabic{figure}} 
\centering
\subfloat[$\hat S_{BdG}$ spin texture. ]{\includegraphics[width=0.3\textwidth]{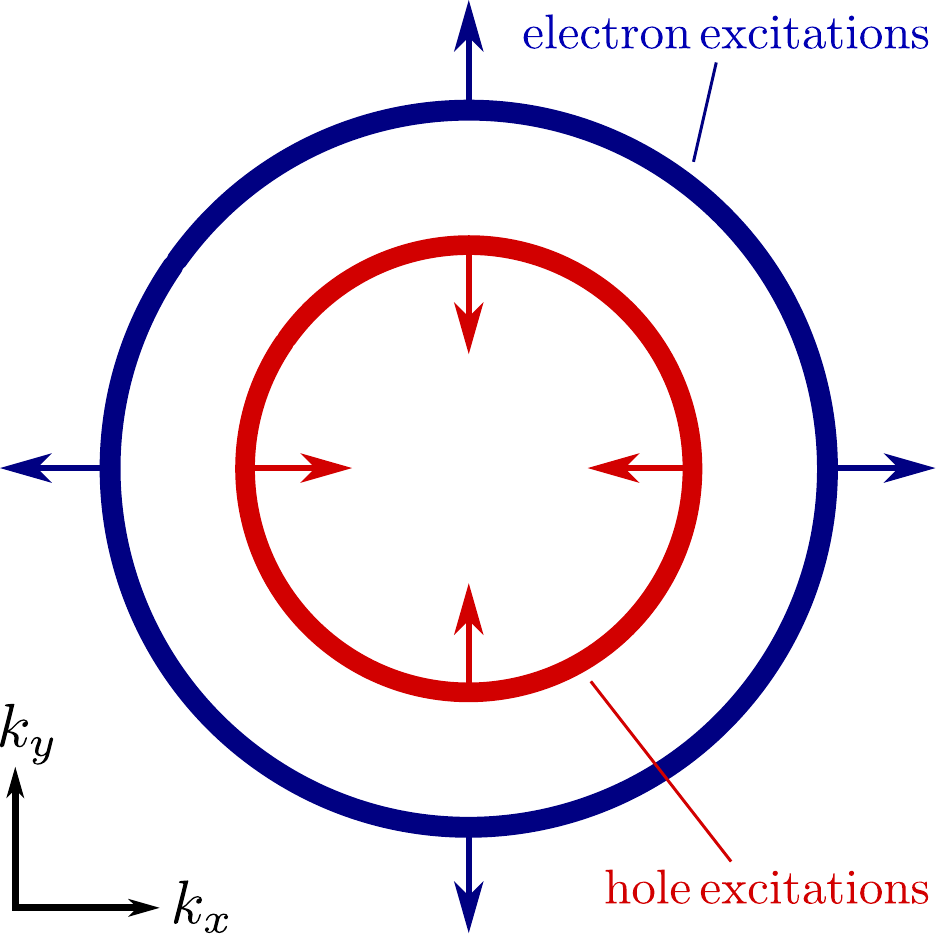}} \hfill
\subfloat[$\hat S_{CP}$ spin texture (local).]{\includegraphics[width=0.3\textwidth]{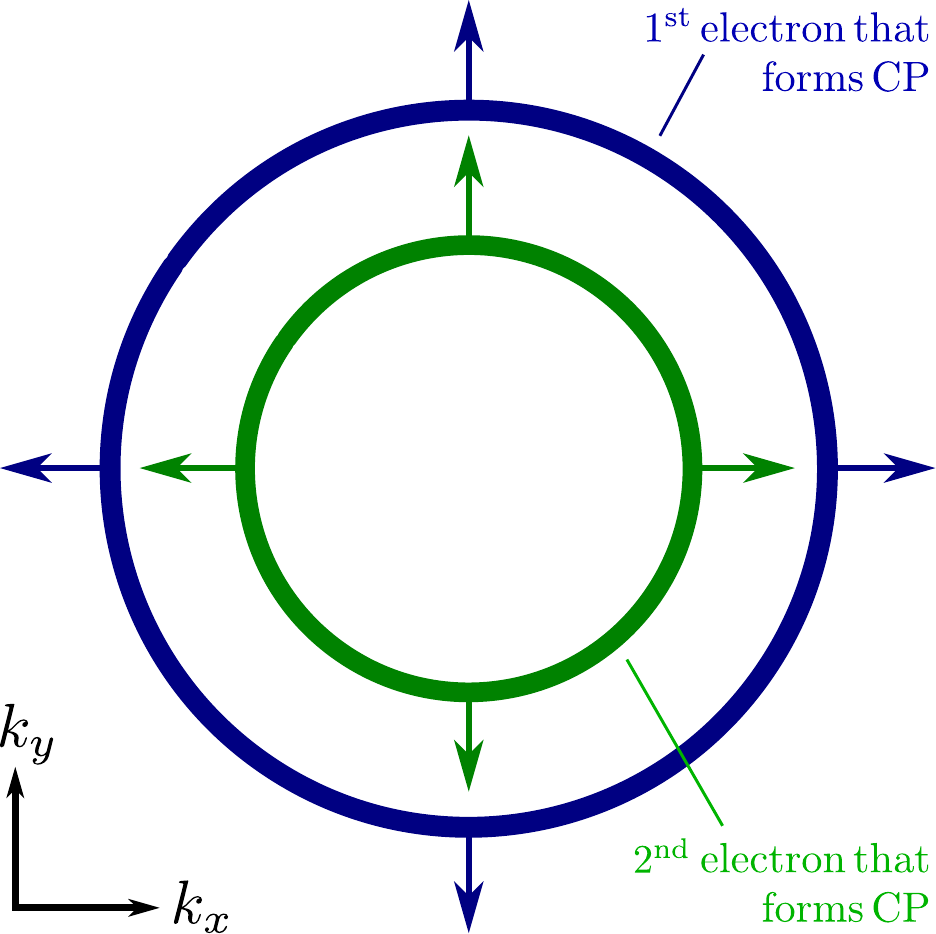}} \hfill
\subfloat[$\hat S_{CP}$ spin texture (non--local).]{\includegraphics[width=0.3\textwidth]{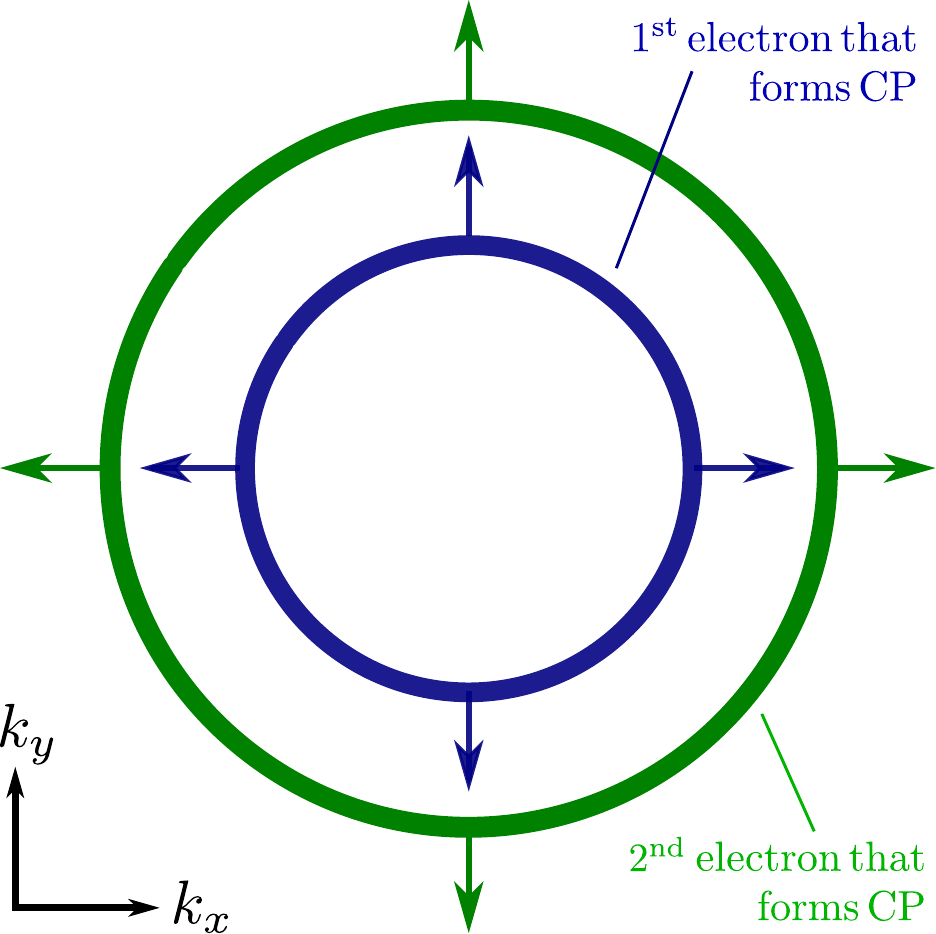}} 
\caption{Spin textures for a constant excitation energy $\varepsilon>-\mu_R>\mu_L$. \textbf{(a)} $\hat S_{BdG}$ gives the spin (indicated by the arrows) of the electron (blue) and hole (red) quasi--particle states described by the BdG formalism. Electrons and holes with the same $\bsy k/\abs{\bsy k}$ feature exactly the opposite spin. In contrast, $\hat S_{CP}$ gives the correct spin of the two electron states that form the CP inside the SC if the incident electron is excited in $L$ and the second electron state (green) is removed from the Fermi sea in \textbf{(b)} $L$ (LAR) or \textbf{(c)} $R$ (CAR). The spin of the removed electron is always opposite to that of the corresponding hole for a given $\bsy k$. Note that in \textbf{(c)}, we plot over a larger range; the total electron momentum $\bsy k$ shall be the same in all three sketches.} 
\label{fig:A3}
\end{figure}

\begin{figure}[ht]
\renewcommand{\figurename}{Fig.}
\renewcommand\thefigure{A\arabic{figure}} 
\centering
\subfloat[{$\mathrm{e} V_L \in [0,0.6\Delta_0 ].$}]{\includegraphics[width=0.48\textwidth]{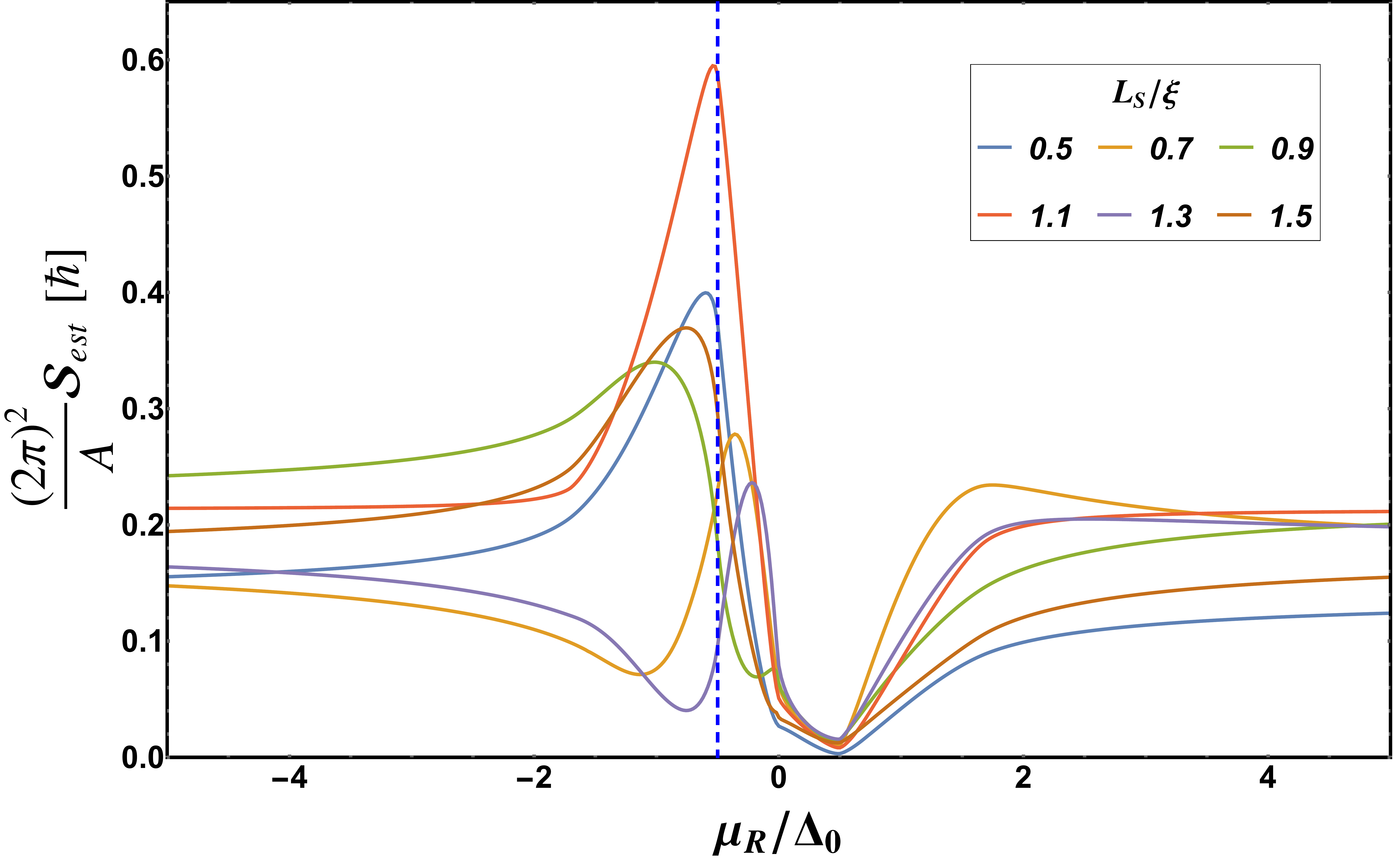}} \hfill
\subfloat[{$\mathrm{e} V_L \in [0,\Delta_0 ].$}]{\includegraphics[width=0.48\textwidth]{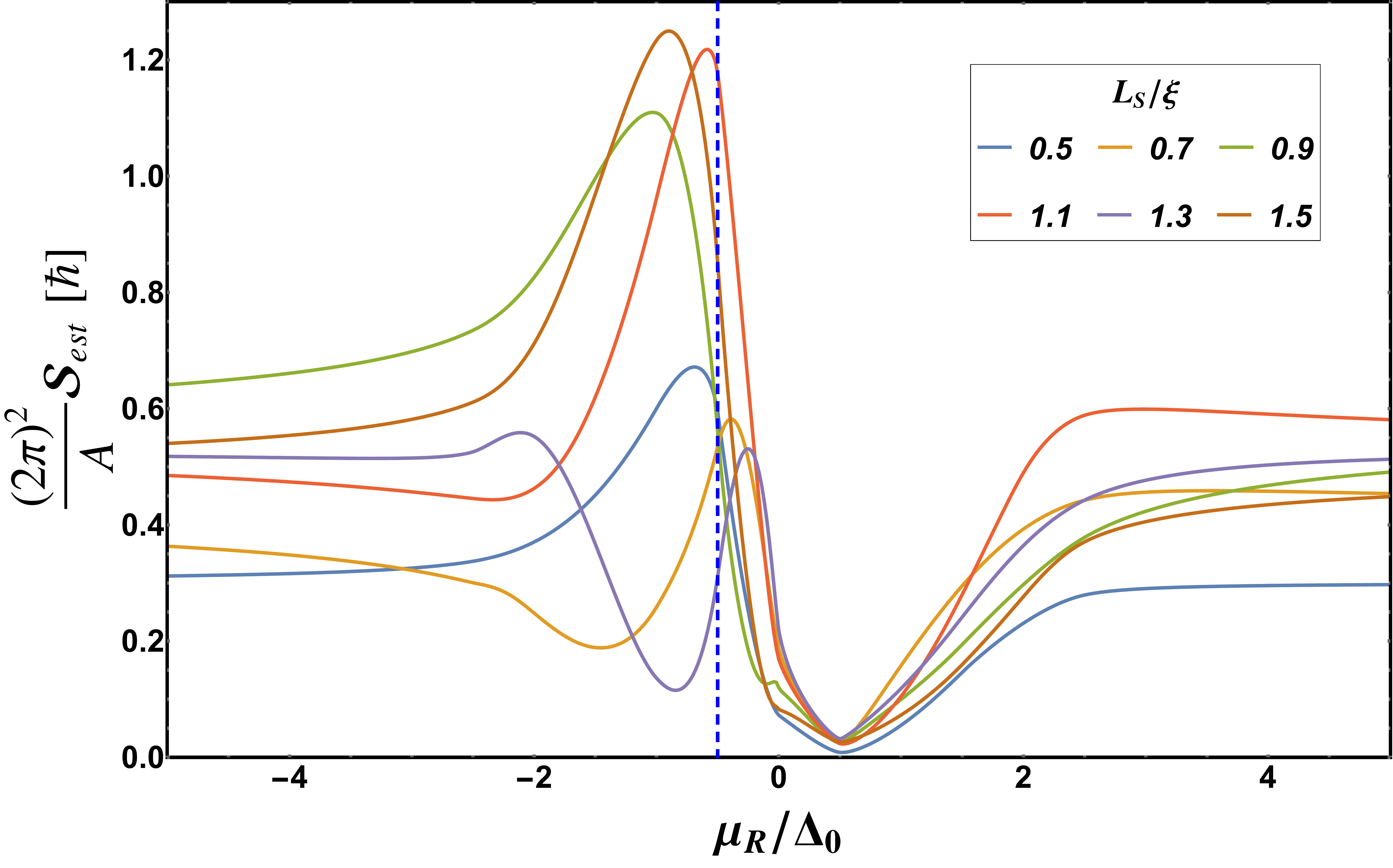}
} 
\caption{Estimated net spin pumped into the SC as a function of the electrochemical potential $\mu_R$. For different lengths of the SC, we plot the results for a bias increased from $\mathrm e V_L=0$ to \textbf{(a)} $\mathrm e V_L=0.6\Delta_0$ and \textbf{(b)} $\mathrm e V_L=\Delta_0$. We choose $\mu=0.5\Delta_0$ and $\mu_S=10\Delta_0$ in the figure.  }
\label{fig:A2}
\end{figure}

\noindent In our basis, $\left(\hat c_\uparrow(\bsy k),\hat c_\downarrow(\bsy k),\hat c^\dagger_\uparrow(-\bsy k),\hat c^\dagger_\downarrow(-\bsy k)\right)^T$, the operator to calculate the spin of the BdG quasi--particles is given by 
\begin{align}
\label{A14a}
\hat S_{BdG}=\frac \hbar 2
\begin{pmatrix}
\vec{\bsy \sigma} & 0 \\ 0& -\vec{\bsy \sigma}^*
\end{pmatrix},
\end{align}
with the vector of Pauli matrices $\vec{\bsy \sigma}=\left(\p x,\p y,\p z \right)$. For a constant excitation energy $\varepsilon>\mu_L$, we obtain the electron and hole spin texture in region $L$ as illustrated in Fig. \ref{fig:A3} \textbf{(a)}. An electron and a hole with the same orientation of the momentum, $\bsy k/\abs{\bsy k}$, have the exact opposite spin in this picture. For LAR, with $k_y$ conserved, this means that the spins of the incident electron and the reflected hole point in a similar, but not exactly the same direction (except for $\varepsilon=0$, where they are parallel). \\
\indent The spin of the reflected hole, however, is not identical to that of the electron removed from the Fermi sea to form the Cooper pair (CP) in the SC. We argue that, since the Fermi sea can be assumed to have zero angular momentum before the electron was removed, the hole described by the BdG formalism has the exact opposite spin of this removed state. Hence, $\hat S_{BdG}$ is not the appropriate operator to calculate the total spin of the CP. Instead, we introduce what we denote Cooper pair spin operator, as follows
\begin{align}
\label{A14b}
\hat S_{CP}=\frac \hbar 2
\begin{pmatrix}
\vec{\bsy \sigma} & 0 \\ 0& \vec{\bsy \sigma}^*
\end{pmatrix}.
\end{align}
With this operator, we obtain the spin textures of the incident and removed electron in a LAR process shown in Fig. \ref{fig:A3} \textbf{(b)}. While the spin of the incident electron in a given state does not change, that of the Fermi sea electron is opposite to the spin of the corresponding Andreev reflected hole in the BdG Hamiltonian. Note that, in this picture, the incident electron has positive $x$--momentum ($k_x>0$), while that of the reflected hole is negative ($k_x<0$). 
\indent  We plot the spin texture for the non--local CAR process (we choose the same $\varepsilon>-\mu_R>\mu_L$ as above) in Fig. \ref{fig:A3} \textbf{(c)}. While, for this choice, the hole states in $R$ have a larger total momentum $\bsy k$ than the electrons in $L$, the spin features do not change for a given $\bsy k/\abs{\bsy k}$. The difference now is that, in the CAR process, both the incident electron and the transmitted hole (and with this, the removed electron) have positive $x$--momentum, $k_x>0$. \\
\indent Following the argumentation above, we obtain the correct CP spin (and with this, the net spin pumped into the SC) by evaluation of the expectation values of the operator $\hat S_{CP}$ as defined in Eq. \eqref{A14b}. For convenience, we drop the index notation in the main text and write $\hat S_{CP}\hat = \hat S$.
\FloatBarrier
We want to give an estimate on the effect of the spin pumped into the SC in a certain bias range $\delta \mathrm eV_L=\delta \varepsilon \; (\varepsilon>0)$ by summing the net spin over all modes $k_x$ and $k_y$ in this domain,
\begin{align}
\label{eq:A15}
\bsy{\mathcal{S}}_{est}=\sum\limits_{k_x,k_y}\bsy{\mathcal{S}}(k_x,k_y).
\end{align}
To go to the continuous limit, we choose $\delta k_x=\frac{2\pi}{L_{Sys}}$ and $\delta k_y=\frac{2\pi}{W}$ with $L_{Sys}$ and $W$ the length and the width of the system, respectively. Also, since $k_x$ is determined by $\varepsilon$, we introduce the quantity $\partial\varepsilon/\partial k_x$ and rewrite the sum in Eq. \eqref{eq:A15} to
\begin{align}
\bsy{\mathcal{S}}_{est}&=\sum\limits_{k_x,k_y}\bsy{\mathcal{S}}(k_x,k_y)\to\frac{A}{(2\pi)^2}\int\int \D k_x\D k_y\,\bsy{\mathcal{S}}(k_x,k_y)=\frac{A}{(2\pi)^2}\int\int \D\varepsilon\D k_y \left(\frac{\partial \varepsilon}{\partial k_x}\right)^{-1}\,\bsy{\mathcal{S}}(k_x(\varepsilon),k_y),
\end{align} 
with the area $A=L_{Sys}\cdot W$. Results for different lengths of the SC are given in Fig. \ref{fig:A2}.
For a bias increased from zero to slightly above $\mu_L$, see Fig. \ref{fig:A2} \textbf{(a)}, we find a distinct peak for $L_S\approx1.1\xi$ in the bipolar setup. Increasing the bias further, see Fig. \ref{fig:A2} \textbf{(b)}, the system features comparable effects for different lengths of the SC (here, e.g., for $L_S=0.9\xi$ and $L_S=1.5\xi$) at different electrochemical potentials $\mu_R$ in $R$. However, in contrast to the bipolar setup, for these choices the non--equilibrium singlet CPs are not completely filtered out.
\end{document}